\RequirePackage{fix-cm}
\documentclass[smallextended]{svjour3}       % onecolumn (second format)
\smartqed  % flush right qed marks, e.g. at end of proof
\usepackage{graphicx}
\usepackage[usenames,dvipsnames]{xcolor}
\usepackage[normalem]{ulem}
\begin{document}

\title{Electromagnetic field and cosmic censorship%\thanks{Grants or other notes
}


\author{Koray D\"{u}zta\c{s}       %\and
        %Second Author %etc.
}


\institute{K. D\"{u}zta\c{s}  \at
            Bo\u{g}azi\c{c}i University, Department of Physics \\ Bebek 34342, \.Istanbul, Turkey \\
              \email{koray.duztas@boun.edu.tr}           %  \\
           %\and
          % S. Author \at
             % second address
}

\date{Received: date / Accepted: date}

\maketitle

\begin{abstract}
We construct a gedanken experiment in which an extremal Kerr black hole interacts with a test electromagnetic field. Using Teukolsky's solutions for electromagnetic perturbations in Kerr spacetime, and the conservation laws imposed by the energy momentum tensor of the electromagnetic field and the Killing vectors of the spacetime, we prove that this interaction cannot convert the black hole into a naked singularity, thus cosmic censorship conjecture is not violated in this case.
\end{abstract}

\section{Introduction}
\label{intro}
The original form of the cosmic censorship conjecture \cite{penrose.orig.ccc} (CCC) --which was later called the \emph{weak} CCC as opposed to the strong one \cite{penrose.strong.ccc}-- asserts that ``naked singularities" can not evolve starting from nonsingular initial data. For reviews see  \cite{ccc.rev.1,ccc.rev.2,ccc.rev.3}.

Naked singularities are those that are not hidden behind an event horizon, i.e. they have access to the asymptotically distant regions of the universe. This access would make the specification of a well defined initial value problem impossible, since initial conditions can not be specified at a singularity. That is why naked singularities are conjectured not to exist, or at least not to evolve from regular initial conditions.

A general proof of the CCC has proved elusive. Hence, one had to meanwhile settle for looking for weak spots in the weak version of the conjecture; i.e. imagining initial conditions that are regular yet seem destined to evolve into a naked singularity, and analytically or numerically analyzing the system to see if it really does. Given that an event horizon is the defining feature of a black hole, it is very natural to ask that question in their context. 

Another feature that facilitates the testing of WCCC in the black hole context is the no-hair theorem \cite{mazur}  in classical general relativity which states that stationary, asymptotically flat spacetimes are uniquely parametrized by three parameters, (Mass $M$, charge $Q$, and angular momentum per unit mass $a$) and the existence of the horizon depends on the validity of an inequality between these parameters, namely 
\begin{equation}
M^{2} \geq Q^{2}+a^{2}. \label {criterion}
\end{equation}
in appropriate units. In other words, a spacetime described by the Kerr-Newman metric corresponds to a black hole, if (\ref{criterion}) is satisfied; but to a naked singularity if it is violated. Therefore in this context the question is if it is possible to manipulate the Kerr-Newman parameters of a spacetime so that a black hole satisfying (\ref{criterion}) evolves to a naked singularity. This manipulation can be envisaged as the black hole absorbing some particles or fields coming from infinity. The no-hair theorem guarantees that once the particles/fields are absorbed/reflected, the space-time will settle to another Kerr-Newman space-time, provided that the system settles down to a stationary configuration~\cite{Jacobson-Sot0}. The modification of the Kerr-Newman parameters can be calculated reasonably easily only in the test particle/field approximation, i.e. when the energy momentum etc. of the particle/field makes negligible local impact on the geometry. Hence, in the calculable case, the change in the Kerr-Newman parameters is infinitesimal, therefore we must start our thought experiment from conditions infinitesimally close to where we would like to push the system, which corresponds to the case where the inequality (\ref{criterion}) is saturated  --the so-called ``extremal" black hole--. 

The first experiment in this vein was constructed by Wald \cite{wald74}. He showed that particles with enough charge and/or angular momentum to overcharge or overspin a black hole either miss, or are repelled by, the black hole. This result was generalised to the case of dyonic black holes for spinless test particles \cite{hiscock} and charged massive scalar test fields \cite{dkn}. A review of related work \cite{bekenst-rosenzwg,hod,hubeny,hod.3,Jacobson-Sot,deFelice-yu} about thought experiments attempting to overcharge/overspin extremal or nearly extremal black holes, up to approximately 2009, is given in the introduction of \cite{dkn}; we may also add here  \cite{bck,rs,ist,toth}. There are also works extending the test-particle result to slightly subextremal Kerr-Newman black holes~\cite{saa_santarelli}, and suggesting that quantum tunnelling can lead to violation of WCCC~\cite{matsas_daSilva,ri_saa_1,matsasEtal,ri_saa_2}. These works seem to ignore quantum radiation and back reaction effects which would reset WCCC~\cite{hod2,hod4}. WCCC was also challenged with spherical shells~\cite{hod3},  its violation was claimed to be possible even for extremal black holes and test particles, due to higher order terms~\cite{gao_zhang_1}, and critical black holes with cosmological constant were analysed in this context~\cite{gao_zhang_2}.

In the present work, we consider Kerr black holes interacting with a free electromagnetic \emph{test} field to see if it is possible for the black hole to turn into a naked singularity; i.e. we continue along the Wald~\cite{wald74} - Hiscock~\cite{hiscock} - Semiz~\cite{dkn} line of investigation. With the same proviso, this work also fills a gap left in \cite{dkn}, where a ``no photons" assumption was made, i.e. the electromagnetic field was considered to be totally sourced by the complex scalar field. Furthermore, for spin-1 fields, we generalize previous work~\cite{overspin}, where Kerr black holes and {\em single-frequency-dominated} fields {\em of spin 0, 1 or 2} were considered. In that work, the WCCC was found to hold for extremal black holes, but seemed to be violable for slightly subextremal black holes, analogously to~\cite{hubeny} and~\cite{Jacobson-Sot}.

The present thought experiment  also involves a packet of electromagnetic waves (spin 1) incident on the black hole from infinity, however the analysis in this work is not constrained to modes of a single (dominant) frequency $\omega$ and a single azimuthal wave number $m$. In this work we evaluate the validity of WCCC for the most general solution of electromagnetic fields in Kerr space-time in the form of a superposition of each mode $(l,m,\omega)$ with arbitrary coefficients $f_{lm}(\omega)$ showing that mode's contribution to the wave packet. The interaction of the field with the black hole results in partial transmission of the field into the black hole and partial reflection back to infinity, i.e. we deal with a scattering problem.  After the field decays away the final state is characterized by another Kerr space-time, with new values of $M$ and $a$. 

The net radial flux of energy and the net radial flux of angular momentum across any sphere centered at the black hole are given by surface integrals of $-T^{r}_{\;\;t}$ and $T^{r}_{\;\;\phi}$ respectively, where $T_{\mu\nu}$ is the energy momentum tensor for the field.  If the contribution of each mode to the fluxes of energy and angular momentum can be calculated independently and subsequently added, the derivation in~\cite{overspin} which is valid for single frequency modes can be directly generalized to the case of superposition of different modes. This is indeed the case for scalar fields (see e.g. \cite{overspin} section 4, or \cite{dkn}). However, for electromagnetic fields it is not obvious that there will be no cross terms when we constitute the flux integrals for a superposition of modes, considering the form of the energy momentum tensor (see section (\ref{section3})), so the derivation of the validity of WCCC for single frequency modes does not necessarily apply to the most general solution for electromagnetic fields in Kerr space-time which is given by a superposition of all modes. In this work we consider the most general solution for electromagnetic fields in Kerr space-time, and construct expressions for the fluxes of energy and angular momentum carried by the electromagnetic field in order to establish the concrete proof for the validity of WCCC when the field is incident on an extremal  Kerr black hole.

We treat the free electromagnetic test field using Newman-Penrose (NP) \cite{newpen} formalism. Source-free Maxwell equations for the relevant NP variables are separated in the Kerr spacetime, and asymptotic solutions at the horizon and at infinity are found by Teukolsky \cite{teuk2}.  The question we ask and answer in this work is, if the Weak Cosmic Censorship Conjecture can be violated in the context described above. 

\section{Teukolsky's results for electromagnetic field in Kerr geometry}
\label{sec:1}
Separable wave equations for electromagnetic and gravitational perturbations of Kerr black holes were studied in detail by Teukolsky \& Press~\cite{teuk2,teukII,teukIII}, where they decouple Maxwell's equations and gravitational equations to combine them into a single master equation and derive separated equations which are ordinary differential equations with simple asymptotic solutions. This section consists of a review of their results. 

The derivation applies to any Type D vacuum background metric (which includes the Kerr solution). Choosing the null vectors $l$ and $n$ of the Newman-Penrose tetrad \cite{newpen} along the repeated principal null directions of the Weyl tensor we have
\begin{equation}
\kappa = \sigma=\nu=\lambda=0 \label{pnull}
\end{equation}
for four of the twelve NP spin coefficients. Then the Maxwell's equations for a test field in electrovacuum are
\begin{eqnarray}
& &(D-2\rho)\phi_1 -(\delta^* +\pi-2\alpha)\phi_0 =0  \label{maxwell1}\\
& &(\delta -2\tau)\phi_1 -(\Delta +\mu -2\gamma)\phi_0 =0 \\
& &(D-\rho +2\epsilon)\phi_2 -(\delta^* +2\pi)\phi_1=0 \\
& &(\delta -\tau +2\beta)\phi_2 -(\Delta +2\mu)\phi_1 =0 \label{maxwell4}
\end{eqnarray}
where $\phi_{0}$, $\phi_{1}$ and $\phi_{2}$ are complex scalars representing the electromagnetic field in the NP formalism; $D$, $\delta$, $\Delta$ and $\delta^{*}$ are NP  differential operators and $\rho$, $\pi$, $\alpha$, $\tau$, $\mu$, $\gamma$, $\epsilon$ and $\beta$, the remaining 8 NP spin coefficients. The NP differential operators are first order and are associated with one member of the tetrad each; and $\phi_{0}$, $\phi_{1}$ and $\phi_{2}$ are related to the electromagnetic field by
\begin{eqnarray}
& &\phi_0=F_{\mu\nu}l^{\mu}m^{\nu} \mbox{ , } \phi_1=\frac{1}{2}F_{\mu\nu}(l^{\mu}n^{\nu}+m^{*\mu}m^{\nu}) \nonumber \\
& & \phi_2=F_{\mu\nu}m^{*\mu}n^{\nu}  \label{defnphi}
\end{eqnarray}
where $m$ is the complex member of the NP tetrad. The relations (\ref{defnphi}) can be inverted to give 
\begin{equation}
F_{\mu\nu}=2[\phi_1(n_{[\mu}l_{\nu]}+m_{[\mu}m^*_{\nu]}) + \phi_2l_{[\mu}m_{\nu]}+ \phi_0 m^*_{[\mu}n_{\nu]}] +\rm{c.c.} \label{fnp}
\end{equation}
From the first order coupled equations (\ref{maxwell1})--(\ref{maxwell4}) one can get decoupled second order equations for $\phi_0$ and $\phi_2$.
\begin{eqnarray}
&&[(D-\epsilon + \epsilon^* -2\rho + \rho^*)(\Delta + \mu - 2\gamma)\nonumber \\ 
&& -(\delta-\beta-\alpha^*-2\tau+\pi^*)(\delta^* +\pi -2\alpha)]\phi_0 =0 \\
&&[(\Delta+\gamma -\gamma^* + 2\mu +\mu^*)(D-\rho +2\epsilon)\nonumber \\ 
&& -(\delta^* +\alpha+\beta^*+2\pi^ -\tau^*)(\delta^ -\tau +2\beta)]\phi_2 =0
\end{eqnarray}
An analogous decoupled equation can also be found for $\phi_1$, but it is not separable in Kerr geometry. However, $\phi_0$ and $\phi_2$ contain complete information about all nontrivial features of the test field \cite{fackipser}. After all, the free electromagnetic field has only two degrees of freedom.

Teukolsky writes out the equations in Kerr metric using a NP tetrad of the form:
\begin{eqnarray}
l^\mu & = & [(r^2 + a^2)/\Delta ,1,0,a/\Delta],\nonumber \\
n^\mu & = & [(r^2+a^2), -\Delta ,0,a]/(2\Sigma) \nonumber \\
m^\mu & =& [ia \sin \theta ,0,1, i/\sin \theta]/[\sqrt{2}(r+ia \cos \theta)]
\label{tetrad0}
\end{eqnarray}
where $\Sigma=r^2+a^2\cos^2\theta$ and $\Delta=r^2-2Mr+a^2$, which should not be confused with the NP derivative operator $\Delta$.

 Teukolsky's master equation ( $4.7$ in \cite{teuk2}) is satisfied by $\phi_0$ for $s=1$ and by $\rho^{-2} \phi_2$ for $s=-1$, where $\rho$ is one of the spin coefficients mentioned above, and for the Kerr metric\footnote{However, all the relations starting with eq. (\ref{pnull}), including the separation and the asymptotic solutions, are meaningful/valid for the Kerr-Newman metric (which is also of Petrov type D) as well, since the latter is obtained from Kerr by replacing $\Delta$ with $r^2-2Mr+a^2+Q^{2}$ (where $Q^{2}$ can be $Q_{e}^{2}+Q_{m}^{2}$ if the black hole has electric charge  $Q_{e}$ and magnetic charge $Q_{m}$); in other words, $Q$ does not appear outside $\Delta$, including the derivatives of $\Delta$. Hence it does not appear in the  derivatives of the metric, where it might have other contribution to the field equations. On the other hand, the electromagnetic field carries no charge, hence the free field will not couple to the field of the black hole, so no $Q$ terms will come from there either.} and the chosen NP tetrad takes the form $\rho=(-1)/(r-ia\cos \theta)$. It should not be confused with the possibly more common name $\rho^{2}$ for the quantity $r^2+a^2\cos^2\theta$, called $\Sigma$ by Teukolsky and in the present work. The master equation can be separated in the form
\begin{equation}
\psi= e^{-i\omega t}e^{im\varphi}S(\theta)R(r) \label{sep}
\end{equation}
where the functions $S(\theta)$ are labeled by $l$, and are complete and orthogonal for given $s$, $m$  and $\omega$, hence an arbitrary solution can be expanded in terms of modes given in eq.(\ref{sep}).

The asymptotic solutions at infinity for such a mode are:
\begin{eqnarray}
\phi_2 &\sim & e^{-i\omega t }e^{im\phi}({}_{-1}S_{lm})(Z_{\rm{in}} e^{-i\omega r*}/r^3 + Z_{\rm{out}} e^{i\omega r*}/r)\nonumber \\
\phi_0 &\sim &  e^{-i\omega t }e^{im\phi}({}_{1}S_{lm})(Y_{\rm{in}} e^{-i\omega r*}/r + Y_{\rm{out}} e^{i\omega r*}/r^3)\nonumber \\
& &
\label{solution0}
\end{eqnarray}
where we have adopted the notation of Teukolsky and Press \cite{teukII,teukIII}; $Y_{\rm{in}}$ and $Z_{\rm{out}}$ are the normalizations of the ingoing and outgoing waves at infinity, and $r^*$ is the tortoise coordinate defined by $dr^*/dr=(r^2+a^2)/\Delta$, so that $r^* \to -\infty$ as the horizon is approached. In this paper, we choose to always enclose the function $S$ in parantheses, so that it is unambigiously clear which function the spin index (1 or -1) belongs to. Note that $\phi_0$ and $\phi_2$ are dominant for ingoing and outgoing waves, respectively. The normalizations are related to each other by \cite{teukIII} 
\begin{equation}
-2\omega^2 Y_{\rm{out}}=BZ_{\rm{out}} \label{yout}
\end{equation}
and
\begin{equation}
B Y_{\rm{in}}=-8\omega^2Z_{\rm{in}} \label{yin}
\end{equation}
where $B$ is a positive constant~\cite{teukIII}. The asymptotic solutions for the radial parts of $\phi_{0}$ and $\phi_{2}$ for modes with given $s$, $m$, $l$ and $\omega$ near the horizon are:
\begin{eqnarray}
& &R\sim  e^{ikr*}  \mbox{ (outgoing) }\\
& &R\sim \Delta^{-s} e^{-ikr*} \mbox{ (ingoing) }
\label{hori0}
\end{eqnarray}
where $k=\omega-m\Omega$, $\Omega=a/2Mr_+$ is the rotational frequency of the black hole. Only the ingoing solutions are physical at the horizon \cite{teukIII}. Therefore the asymptotic form of wave modes near the horizon can be expressed in the form
\begin{eqnarray}
\phi_2 &\sim &\rho^2 Z_{\rm{hole}}\Delta  e^{-ikr*}e^{-i\omega t }e^{im\phi}({}_{-1}S_{lm}) \nonumber \\
\phi_0 &\sim &Y_{\rm{hole}}(\Delta^{-1}) e^{-ikr*} e^{-i\omega t }e^{im\phi}({}_{1}S_{lm}).
\label{solutionhor}
\end{eqnarray}
The normalizations $Y_{\rm{hole}}$ and $Z_{\rm{hole}}$ are related by \cite{teukIII}
\begin{eqnarray}
BY_{\rm{hole}}&=&-32ikM^2r_+^2(-ik +2\epsilon)Z_{\rm{hole}}  \nonumber \\
 &=& 16kMr_+[-2kMr_+-i(r_+-M)] Z_{\rm{hole}}  \label{yzhole}
\end{eqnarray}
where $\epsilon=(M^2-a^2)^{1/2}/4Mr_+$ is defined in \cite{teukIII}.

\section{Testing the cosmic censorship conjecture}
\label{section3}
To check the validity of the WCCC, we need to evaluate the changes in the mass and angular momentum of the black hole due to its interaction with the test fields/particles. As is well-known, the stationary and axisymmetric nature of the Kerr spacetime, that is, the existence of the Killing vectors $\partial /\partial t$ and $\partial / \partial \phi$ for the Kerr metric, allows definition/identification of globally conserved energy and angular momentum for the fields/particles in this spacetime. Therefore the rates of change in the corresponding black hole parameters can be expresses as fluxes into the black hole. The current conservation equation $\nabla_a (T^{ac}K_c)=0$ (where $K$ is a Killing vector) necessary for this correspondence is derived by combining the expression  of the spacetime symmetry in terms of Lie derivatives, $\mathcal{L}_K g=0$ (where $\mathcal{L}$ denotes the Lie derivative), equivalently the Killing equation, $\nabla_{(a}K_{b)}=0$, with the local conservation of energy-momentum $\nabla_c T^{ac}=0$ in general relativity, which follows from the (contracted) Bianchi identity. 

So we have
\begin{equation}
\left(\frac{dM}{dt}\right)_{\rm b.h} = - \int_{S_{\infty}} \sqrt{-g} \, T^{1}_{\;\;0} d\theta d\phi
                                                   \label{eq:dm/dt}
\end{equation}
and
\begin{equation}
\left(\frac{dL}{dt}\right)_{\rm b.h}
=   \int_{S_{\infty}} \sqrt{-g} \, T^1_{\;\;3} d\theta d\phi \label{eq:dl/dt}
\end{equation}
where the label b.h. stands for black hole and $S_{\infty}$ is the spherical surface as $r \longrightarrow \infty$; since the mass of a physical system is defined only in the asymptotically flat part of space at infinity, if it exists.

For the test of the WCCC, let us define an indicator
\begin{equation}
CCC =  M^{2} - a^{2}
\end{equation}
Then, using $a=L/M$
\begin{equation}
\delta (CCC)  =   \int \frac{d(CCC)}{dt}  dt  =  \int \frac{2}{M} \left\{ (M^{2}+a^{2}) \frac{dM}{dt} - a \frac{dL}{dt} \right\} dt \label{deltaccc}
\end{equation} 
implying
\begin{equation}
 \frac{d(CCC)}{dt} = \frac{2}{M}\int_{S_{\infty}} \sqrt{-g} [ (M^{2}+a^{2})(- T^{1}_{\;\;0})   - a T^{1}_{\;\;3} ] d \theta \; d \phi \label{eq:dC/dt1}
\end{equation}
We can use the NP tetrad (\ref{tetrad0}) to derive
\begin{equation}
T_{ab}l^al^b \frac{\Delta^2}{4\Sigma^2}-T_{ab}n^an^b=\frac{\Delta}{\Sigma^2}(r^2+a^2)T_{10}+\frac{a\Delta}{\Sigma^2}T_{13} 
\label{lalbnanb0}
\end{equation}
where we recognize that the right-hand-side is almost proportional to the integrand of (\ref{eq:dC/dt1}). But for the initial and final Kerr black holes, no contribution to the mass (energy) and angular momentum exists outside the horizon, so the time integrated fluxes through surfaces at the horizon and at infinity will be equal. Therefore we can evaluate $\delta(CCC)$ at the horizon. For horizons of extremal black holes, the rhs of (\ref{lalbnanb0}) and the integrand of (\ref{eq:dC/dt1}) {\em do} become proportional. 

For electromagnetic fields, the energy momentum tensor in terms of the corresponding NP scalars is given by (see e.g. \cite{teuk2})
\begin{eqnarray}
4\pi T_{\mu\nu}&=&\{\phi_0 \phi_0^* n_\mu n_\nu + 2\phi_1 \phi_1^* [l_{(\mu}n_{\nu)}+ m_{(\mu}m_{\nu)}^*] +\phi_2 \phi_2^*  l_{\mu}l_{\nu}\nonumber \\
& &-4\phi_1 \phi_0^* n_{(\mu}m_{\nu)}-4\phi_2 \phi_1^* [l_{(\mu}m_{\nu)}+2\phi_2 \phi_0^* m_{\mu}m_{\nu} \} \nonumber \\& &+\rm{c.c.}
\label{tmunu}
\end{eqnarray} 
implying
\begin{equation}
T_{ab}l^al^b=(1/2\pi)\vert \phi_0 \vert^2 \;\;\; {\rm and} \;\;\; T_{ab}n^an^b=(1/2\pi)\vert \phi_2 \vert^2
\end{equation}
hence in Kerr spacetime
\begin{equation}
- (r^2+a^2) T^{1}_{\;\;0} - a T^{1}_{\;\;3} = \frac{\Sigma}{2\pi}\left( \vert \phi_0 \vert^2 \frac{\Delta^2}{4\Sigma^2}-\vert \phi_2 \vert^2 \right)
\label{lalbnanb1}
\end{equation}
which means that
we will be able to evaluate $\delta(CCC)$ although we do not have an explicit solution for $\phi_1$.

Since we are going to work at the horizon, let us first write the most general solution in terms of separated modes according to (\ref{solutionhor}):

\begin{eqnarray}
& &\phi_0= \int d\omega f_{lm}(\omega)e^{-i\omega t} \sum_{l,m} e^{im\phi} [{}_{1}S_{lm}(\theta , a\omega)] Y_{\rm{hole}}(\Delta^{-1}) e^{-ikr*} \nonumber \\
& &\phi_2= \int d\omega g_{lm}(\omega)e^{-i\omega t} \sum_{l,m} e^{im\phi} [{}_{-1} S_{lm}(\theta ,a\omega)] \rho^2 Z_{\rm{hole}}\Delta e^{-ikr*} \nonumber \\
& &
\label{YholeZhole}
\end{eqnarray}
where $f_{lm}(\omega)$ and $g_{lm}(\omega)$ are arbitrary coefficients showing the contribution of the mode $(l,m,\omega)$ to the wave packet. Next we define
\begin{eqnarray}
\phi_0 &\equiv & \Delta^{-1} \psi_0 \nonumber \\
\phi_2 &\equiv & \rho^2 \Delta \psi_2 \label{psi}
\end{eqnarray}
which makes $\psi_0$ and $\psi_2$ regular at the horizon. Substituting these expressions in (\ref{lalbnanb1}), near the horizon we have
\begin{equation}
- (r^2+a^2) T^{1}_{\;\;0} - a T^{1}_{\;\;3} = \frac{1}{2\pi\Sigma}\left( \frac{\vert \psi_0 \vert^2}{4}-\Delta^2\vert \psi_2 \vert^2 \right)
\label{lalbnanb2}
\end{equation}
where we have used $\rho^2 \rho^{*2}=1/\Sigma^2$. Now we evaluate (\ref{lalbnanb2}) at the horizon ($\Delta \to 0$) for extremal black holes ($r_+=M$)
\begin{equation}
- (M^2+a^2) T^{1}_{\;\;0} - a T^{1}_{\;\;3} = \frac{\vert \psi_0 \vert^2}{8 \pi\Sigma}  
\label{lalbnanb3}
\end{equation}
The left hand side of (\ref{lalbnanb3}) is the integrand in  (\ref{eq:dC/dt1}). We can evaluate the integral at the horizon:
\begin{equation}
 \frac{d(CCC)'}{dt} = \frac{2}{M}\int_{S_{\rm{H}}} \sqrt{-g} \left[ \frac{\vert \psi_0 \vert^2}{8 \pi\Sigma} \right] d \theta \; d \phi \label{eq:dC/dt2}
\end{equation}
and
\begin{equation}
\delta(CCC) = \frac{2}{M}\int\int_{S_{\rm{H}}} \sqrt{-g} \left[ \frac{\vert \psi_0 \vert^2}{8 \pi\Sigma} \right] d \theta \; d \phi dt \label{cccfin}
\end{equation}
since as argued above, $\delta(CCC)$ can be calculated either at infinity or at the horizon for the Kerr black hole.

The expression (\ref{cccfin}) is strictly positive. Thus, one can not overspin an extremal Kerr black hole by sending in electromagnetic test fields. Note that the modes with $\omega<m \Omega$, where $\Omega=a/(r_+^2+a^2)$ is the angular velocity of the horizon, carry more angular momentum than energy. The absorption of these modes is expected to give a negative value of $\delta (CCC)$. The positive definiteness of $\delta (CCC)$ indicates that there is no net absorption of the modes in the range $0<\omega<m \Omega$. This is in accord with the well-known effect of {\it superradiance} which occurs for integer-spin fields: For $0<\omega<m\Omega$, the incident waves are amplified as they scatter off the black hole, and not absorbed by it \cite{misner,teuknature}. 

%\subsection{Subsection title}
%\label{sec:2}
%as required. Don't forget to give each section
%and subsection a unique label (see Sect.~\ref{sec:1}).
%\paragraph{Paragraph headings} Use paragraph headings as needed.
%\begin{equation}
%\end{equation}

%\begin{figure}
 % \includegraphics{example.eps}
%\end{figure}
%\begin{figure*}
 % \includegraphics[width=0.75\textwidth]{example.eps}
%\end{figure*}
%\begin{table}
%\begin{tabular}{lll}
%\hline\noalign{\smallskip}
%\end{tabular}
%\end{table}

\begin{acknowledgements}
I would like to thank {\. I}  Semiz for helpful discussions and comments.
\end{acknowledgements}



\end{document}